\documentclass[12pt,twoside]{article}
\usepackage{amsfonts,graphicx}

\newtheorem{theorem}{Theorem}
\newtheorem{lemma}{Lemma}
\newtheorem{proposition}{Proposition}
\newtheorem{corollary}{Corollary}
\newtheorem{definition}{Definition}
\newtheorem{remark}{Remark}

\newcommand{\R} {{\mathbb R}}
\newcommand{\Q} {{\mathbb Q}}
\newcommand{\Z} {{\mathbb Z}}
\newcommand{\N} {{\mathbb N}}

\newcommand{\bi} {billiard}
\newcommand{\me} {measure}

\newcommand{\tr} {trajector}
\newcommand{\eo} {escape orbit}
\newcommand{\erg} {ergodic}
\newcommand{\eps} {\varepsilon}
\newcommand{\st} {\: \big| \:}
\newcommand{\fat} {\phi_{\vartheta,t}}
\newcommand{\ft} {\phi_{t}}
\newcommand{\qed} {\hfill {\small Q.E.D.} \par\medskip}
\newcommand{\skippar} {\par\medskip}

\newcommand{\proof} {\textsc{Proof.} }
\newcommand{\proofof}[1] {\textsc{Proof of {#1}.} }
\newcommand{\article}[3] {\textsc{{#1}}, {\itshape {#2}}, {{#3}}.}
\newcommand{\book}[3] {\textsc{{#1}}, {\itshape {#2}}, {{#3}}.}
\newcommand{\vol} {\textbf}

\newcommand{\Bi} {P}
\newcommand{\Bin} {\Bi^{(n)}}
\newcommand{\Su} {S}
\newcommand{\s} {{\cal S}}
\newcommand{\Pm} {{\cal P}}

\renewcommand{\iff} {if, and only if,\ }
\renewcommand{\o} {orbit}
\renewcommand{\tilde} {\widetilde}
\renewcommand{\a} {\alpha}

\begin{document}

\title{
	Escape Orbits and Ergodicity in Infinite Step Billiards
}

\author{ 
	Mirko Degli Esposti $^{a,b)}$, Gianluigi Del Magno $^{b)}$,
	Marco Lenci $^{c)}$ \\ 
	\\ 
	$^{a)}$ Dipartimento di Matematica \\ 
	Universit\`a di Bologna \\ 
	40127 Bologna, Italy \\ 
	\\
	$^{b)}$ School of Mathematics \\ 
	Georgia Institute of Technology \\ 
	Atlanta, GA \ 30332, U.S.A. \\ 
	\\ 
	$^{c)}$ Mathematics Department \\ 
	Rutgers University \\ 
	Piscataway, NJ \ 08854, U.S.A. \\ 
	\\ 
	{\footnotesize E-mail: \ttfamily desposti@dm.unibo.it, 
	magno@math.gatech.edu, lenci@math.rutgers.edu} 
}

\date{June 1999}

\maketitle

\begin{abstract}
	In \cite{ddl} we defined a class of non-compact polygonal \bi
	s, the \emph{infinite step \bi s}: to a given sequence of
	non-negative numbers $\{ p_{n} \}_{n\in\N}$, such that $p_{n}
	\searrow 0$, there corresponds a \emph{table} $\Bi :=
	\bigcup_{n\in\N} [n,n+1] \times [0,p_{n}]$.

	In this article, first we generalize the main result of
	\cite{ddl} to a wider class of examples. That is, a.s.~there
	is a unique \emph{\eo} which belongs to the $\alpha$- and
	$\omega$-limit of every other \tr y. Then, following the
	recent work of Troubetzkoy \cite{tr}, we prove that
	\emph{generically} these systems are \erg\ for almost all
	initial velocities, and the entropy with respect to a wide
	class of \erg\ \me s is zero.
	
	\medskip\medskip
	\noindent
	AMS classification scheme number: 58F11
\end{abstract}

\section{Introduction}\label{intro}

\begin{figure}[ht]
   
   \hspace{4.5cm}	
   \includegraphics[width=2in]{fig0.eps}
   \protect\label{omino}
\end{figure}

Playing on a rectangular billiard with just one ball can be a boring
experience, in particular when the ball is represented by a point:
given any starting position, the corresponding orbit is dense on the
billiard for almost all directions, whereas for a countable set of
initial angles all orbits are periodic.

We can make the game a little more interesting if we use a generic 
polygonal \bi.

\begin{remark} 
	In more generality, billiards are dynamical systems defined by
	the uniform motion of a point inside a
	piecewise-differentiable planar domain \emph{(the table)},
	with elastic reflections at the boundary. This means that the
	tangential component of the velocity remains constant and the
	normal component changes sign \cite{t}.
\end{remark}

A polygonal table is called \emph{rational} when the angles between
the sides are all of the form $\pi n_{i}/m_{i}$, where $n_{i}$ and
$m_{i}$ are integers.  These billiards have the nice property that any
orbit will have only a finite number of different angles of reflection
$\vartheta$. In this case, there is a natural \emph{unfolding}
procedure \cite{fk,kz} which allows one to map any given trajectory on
the table to an orbit of a vector field over a surface having a certain
genus.

\begin{remark} \label{remark2}
	Usually one assumes that the magnitude of the particle's
	velocity equals one, and that the \o\ which hits a vertex
	stops there. For our model, though, we will slightly modify
	this second assumption, since it is possible to continue
	uniquely every \tr y for all values of $t$ (see later).
	Either way, the set of initial conditions whose \o s do not
	contain any vertices is always a set of full \me\ in the phase
	space.
\end{remark}

The aforementioned rational condition implies in particular a
decomposition of the phase space in a family of flow-invariant
surfaces $\Su(\vartheta),\ 0 \le \vartheta \le \pi/m, \ m:=
l.c.m.\{m_i\}$.  Excluding the special cases $\vartheta =
0,\pi/m$, it is known that the \bi\ flow restricted to any of the
$\Su(\vartheta)$ is essentially equivalent to a geodesic flow $\fat$
on a closed oriented surface $S$, endowed with a flat Riemannian
metric with \emph{conical} singularities \cite{g3}.

Based on this fundamental idea, many important and deep results have
been proven, with the crucial aid of some highly non-trivial
mathematical techniques. In particular, the theory of quadratic
differentials over Riemann surfaces, the theory of interval exchange
transformations (using the induced map on the boundary) and the method
of approximating irrational \bi s with rational ones, have altogether
turned out to be very effective to understand the dynamics of
polygonal billiards.  References \cite{g2,g3} give a complete account
of the state-of-the-art in this field, and \cite{ddl}, of which this
paper is a follow-up, contains a compact reference list.

Here we limit ourselves to recall only those results that will be of
immediate use in the rest of the paper:

\begin{itemize}

\item[(i)] {\rm \cite{kms,ar,kz}} The Lebesgue \me\ in a (finite)
rational polygon is the unique ergodic \me\ for the \bi\ flow, for
(Lebesgue-)almost all directions. Moreover, for all but countably many
directions, a rational polygonal \bi\ is \emph{minimal} (i.e., all
infinite semi-\o s are dense).  In particular, for \emph{almost
integrable} \bi s, ``minimal directions'' and ``ergodic directions''
coincide \cite{g1,g2,b}. (Roughly, an almost integrable \bi\ is a \bi\
whose table is a finite connected union of pieces belonging to a
tiling of the plane by reflection, e.g, a rectangular tiling, or a
tiling by equilateral triangles, etc.---see \cite{g2,g3} for more
precise statements.)

\item[(ii)] {\rm \cite{kms}} For every $n$, in the space of $n$-gons
there is a dense $G_\delta$-subset of \erg\ tables.

\item[(iii)] {\rm \cite{g3,gkt}} For any given polygon, the metric
entropy with respect to any flow-invariant measure is
zero. Furthermore, the topological entropy is also zero.

\item[(iv)] {\rm \cite{gkt}} Given an arbitrary polygon and an \o,
either the \o\ is periodic or its closure contains at least one
vertex.

\end{itemize}

All of these results, however, fail to hold when we introduce further
complications: an infinite number of sides (an interesting example is
the ``staircase'' compact \bi\ briefly mentioned in \cite{efv}, while
the inspiring paper by Troubetzkoy \cite{tr} discusses the general
case), or non-compact tables (one might take a look at \cite{gu,le},
although they are not about flat polygonal \bi s).  
\skippar

\begin{figure}[ht]
    \hspace{3cm}		
    \includegraphics[width=3in]{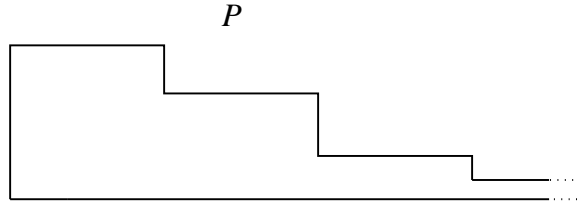}
    \caption{\small An infinite step billiard.}
    \protect\label{fig1}
\end{figure}

In \cite{ddl} we introduced a family of (seemingly) simple models that
have both properties: the \emph{infinite step \bi s}
(Fig.~\ref{fig1}).

\medskip


For them, it turns out that one of the first non-obvious questions to
treat is exactly what our friend that likes playing pool more than
handling groups of automorphisms would like to know: depending on how
I initially hit the ball, can I be sure it will come back at all? In
other words, since the system is \emph{open} it makes sense to ask how
many \o s actually keep returning to the ``bulk'' of the systems
rather than \emph{escaping}. As a matter of fact this was one of our
main motivations in \cite{ddl}. In the one example we treated in
detail there, we found that the fact that there is a unique \eo\ for
a.a. directions has very interesting implications for the dynamics of
the entire billiard: specifically, this \o\ is somehow an
``attractor'' for the system. We will re-state this more precisely in
Section \ref{result-top-dyn}.

Here we consider the whole class of infinite step \bi s and we will
see that playing on these tables can be rather challenging, even if we
pay the price that sometimes we have to wait a \emph{little bit}
before our ball comes back. It turns out that the above result on the
uniqueness of the \eo\ can be extended---together with its
corollaries---to a wide set of \bi s in our class.  For these tables,
the typical situation is that the ball keeps returning to the leftmost
wall, although each time it might stretch arbitrarily far to the
right, following the unique, ill-behaved, escaping \tr y.  This is
basically what Theorem \ref{thm-unique} and related statements
assert---see in particular Proposition \ref{prop-from-ddl}.
(Incidentally, we are pleased to attest that further work in this
direction is in progress \cite{tr2}.)  
\skippar

More theoretically, one would like to know better about the ergodic
properties of these systems (at least of many of them).

Recently Troubetzkoy \cite{tr} treated the case of an arbitrary
polygon with an infinite number of sides, proving interesting results
regarding both the topological structure (e.g., Poincar\'e Recurrence
Theorem and existence of periodic trajectories) and the \erg ity. In
particular, he showed that the \emph{typical} bounded infinite polygon
has ergodic directional flows for almost all angles. The notion of
typicality as first used by \cite{kz} is now customary (see also
\cite{kms}): in a certain topological (usually metric) space there
exists a dense $G_{\delta}$-set of \bi s with that property. The
metric used in \cite{tr} requires several conditions to be verified
before two polygons can be considered close: their shapes are to be
very similar, of course, but also the two flows are to keep close for
a long time, for many initial conditions. (Not that one could do much
better. In fact, it is not hard to imagine two infinite polygons that
look alike but have completely different dynamics---which is exactly
the point in \cite{efv} when the staircase model is presented.)

The other results in this paper apply the ideas of \cite{tr}. We show
that generic \erg ity holds for the family of \emph{unbounded},
finite-volume, billiards in our class (Theorem \ref{gigiTheorem}). The
metric we employ, however, is very simple and can be expressed in
terms of ``easily measurable'' quantities. Also, we prove that the
entropy of the \bi\ is zero w.r.t. any invariant \erg\ \me\ that
verifies a cartain finiteness condition (Proposition \ref{entropy}).

The topological entropy, too, is probably zero (at least for many of
these systems) but, due to the non-compactness of the table, the usual
variational principle cannot be applied directly and one must check
some additional conditions \cite{pp}.  
\skippar

In the next section we introduce the basic notation (referring the
reader to \cite{ddl} for a more detailed presentation of the systems
in question), we give some quick proofs, and we present the precise
statements of our results. The main proofs are then distributed in the
following sections.
\skippar

\textbf{Acknowledgments:} We would like to thank L.~Bunimovich and
S.~Troubetzkoy for stimulating discussions.  M.~D.E. contributed to
this paper during his visit to the School of Mathematics of the
Georgia Institute of Technology, whose support and excellent working
conditions he gratefully acknowledges.

\section{Notations and Statements of the Results} \label{notations}

In this paper we are interested in a class of rational \bi s, the {\em
infinite step \bi s}, defined as follows: Let $\{p_n\}_{n\in\N}$ be a
monotonically vanishing sequence of non-negative numbers, with
$p_{0}=1$ (we will later relax this inessential last condition).  We
denote $\Bi := \bigcup_{n\in\N} [n,n+1] \times [0,p_{n}]$
(Fig.~\ref{fig1}) and we call $(x,y)$ the two coordinates on it.

\begin{figure}[ht]
    \hspace{1cm}	
    \includegraphics[width=4.5in]{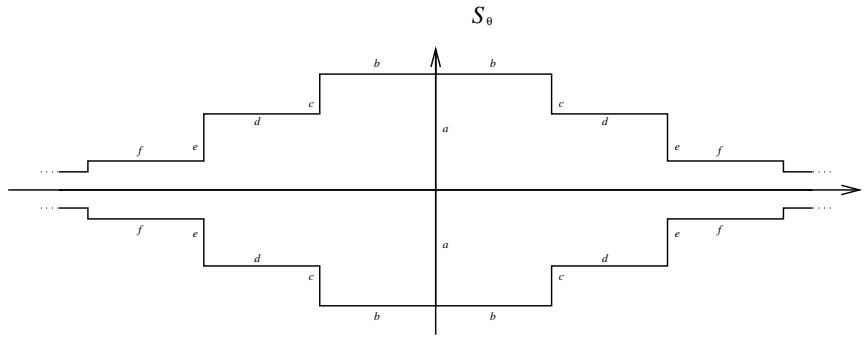}
    \caption{\small The invariant surface $\Su^{\Bi}(\vartheta)$ for the infinite billiard.}
    \protect\label{fig2}
\end{figure}

A point particle can travel within $\Bi$ only in four directions (two
if the motion is vertical or horizontal---cases which we disregard).
One of these directions lies in the first quadrant.  For $\vartheta
\in ]0,\pi/2[$, the invariant surface associated to the billiard flow
is labeled by $\Su^{\Bi}(\vartheta)$ (or just by $\Su(\vartheta)$,
when there is no means of confusion). This manifold is built via the
usual unfolding procedure with four copies of $\Bi$. We denote by
$(X,Y)$ the intrinsic coordinates on $\Su^{\Bi}(\vartheta)$, inherited
by its represention on a plane $(x,y)$, with the proper side
identification (see Fig.~\ref{fig2}).  The $3\pi/2$ corners represent
the \emph{non-removable singularities}, or \emph{singular vertices},
$V_{k}$, of coordinates $(k,p_{k})$ on $\Bi$, or $(\pm k,\pm p_{k})$
on $\Su$.

\medskip\medskip
\medskip\medskip


With the additional condition $\sum_n p_n < \infty$,
$\Su^{\Bi}(\vartheta)$ can be considered a non-compact, finite-area
surface of infinite genus.

\begin{remark}
	In \cite{ddl} we used $R_{\alpha}$ to denote the invariant
	surface for a given $\alpha=\tan\vartheta$. More importantly,
	we measured the directions w.r.t.~the Lebesgue \me\ for
	$\alpha \in ]0,+\infty[$.  However, this is equivalent to the
	normalized Lebesgue \me\ for $\vartheta \in ]0,\pi/2[$. In
	this paper we use angles except in Section \ref{escape}, where
	it is convenient to do otherwise.
\end{remark}

We will denote by $\Bin$ the truncated \bi\ that one obtains by
closing the table at $x=n$.  The corresponding invariant surface, of
genus $n$, will be denoted by $\Su^{(n)}(\vartheta)$
(Fig.~\ref{fig3}). Some statements about $\Bin$ have immediate
consequences for the infinite table $\Bi$. For instance:

\begin{figure}[ht]
    \hspace{1.65cm}
    \includegraphics[width=4.0in]{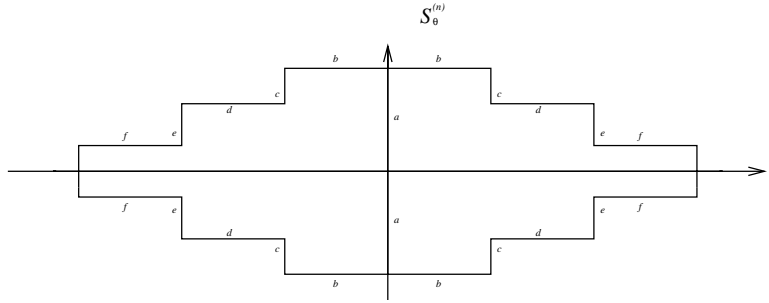}
    \caption{\small The invariant surface $\Su^{(n)}(\vartheta)$ for the truncated
    billiard.}
    \protect\label{fig3}
\end{figure}
\medskip\medskip


\begin{proposition}
	In an infinite step \bi\ $\Bi$, for almost all $\vartheta$,
	all semi-\o s are unbounded, whereas periodic and unbounded
	trajectories can coexist for a zero-\me\ set of directions.
	\label{prop-unbounded}
\end{proposition}

\proof (See also \cite{ddl}, Proposition 4.)  For any given $\vartheta$
and $n>1$, we denote by $\fat^{(n)}$ the flow on
$\Su^{\Bi^{(n)}}(\vartheta)$. Let
\begin{displaymath}
	{\cal M}_n := \{ \vartheta \in ]0,2\pi[ \st \fat^{(n)} 
	\mbox{ is minimal} \}. 
\end{displaymath}
We know that for all $n>1$, $\vert{\cal M}_n\vert=1$ \cite{kms}. Let
${\cal M}_{\infty}=\cap_{n>1}{\cal M}_n$.  Clearly $\vert{\cal
M}_{\infty}\vert =1$. It is easy to see that, for all $\vartheta \in
{\cal M}_{\infty}$, every semi-orbit is unbounded.

On the other hand, in the main example of \cite{ddl} ($p_n = 2^{-n}$),
the truncated tables were almost integrable. Hence $\tan {\cal
M}_\infty^c =\Q \cap \R^{+}$ (the superscript $^{c}$ denoting the
complementary set w.r.t.~a certain space). We found that we had one or
two \eo s for every angle. So, for some \emph{rational directions}
($\alpha\in\Q$), periodic \o s coexisted with \eo s, which are
obviously unbounded. Less trivially, given any arbitrary rational
direction, we showed (in Sec.~1.2) how to construct a table with $p_n
\in \Q$ and with at least one \emph{oscillating} \tr y (this means
unbounded and non-escaping, according to the terminology of
\cite{l}). Playing with that example, it is not hard to show that we
can indeed find a \bi\ with at least one periodic and one oscillating \o.
\qed

\subsection{Topological Dynamics} \label{result-top-dyn}

Call $L :=\{0\} \times [-1,1[$ the closed curve on $\Su^{\Bi}
(\vartheta)$ corresponding to the first vertical side of $\Bi$ (in
Fig.~\ref{fig2} $(0,-1)$ and $(0,1)$ are identified). We will also
occasionally identify $L$ with the interval $[-1,1[$.  The \bi\ flow
along a direction $\vartheta$, which we denote by $\fat$ (or simply by
$\ft$), induces a.e. on $L$ a Poincar\'e map $\Pm_{\vartheta}$ that
preserves the Lebesgue \me. We call it the \emph{(first) return
map}. $\Pm_{\vartheta}$ is easily seen to be an infinite-partition
\emph{interval exchange transformation} (i.e.t.).  On $L$ we establish
the convention that the map is continuous from above: this corresponds
to partitioning $L \simeq [-1,1[$ into right-open subintervals. (More
details in \cite{ddl}, Sec.~1.2---see also Remark \ref{remark2}).

\begin{figure}[ht]
    \hspace{2cm}
    \includegraphics[width=4.0in]{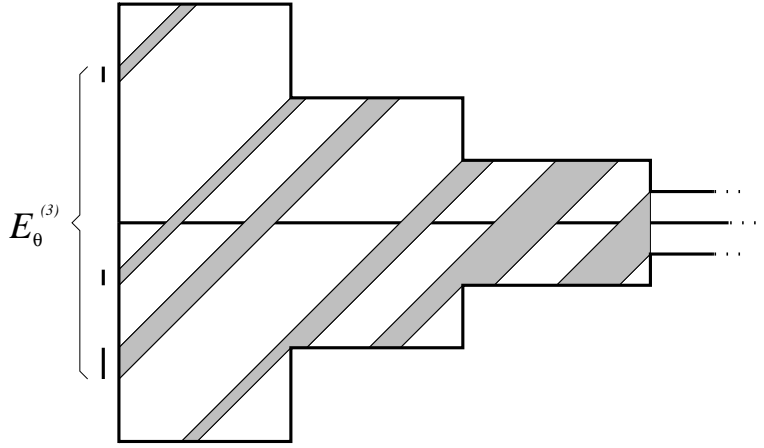}
   
    \caption{\small Construction of $E_{\vartheta}^{(n)}$ as the
 backward evolution of the ``aperture'' $G_{n}$. The beam of orbits
may split at singular vertices.}  
    \protect\label{fig4}
\end{figure}

For a given $\vartheta$ and $n>1$, $G_{n} := \{ n \} \times
[-p_{n},p_{n}[$ denotes the $n$-th \emph{aperture} and
$E_{\vartheta}^{(n)} \subset L$ is the set of points whose forward \o\
starts along the direction $\vartheta$ and reaches $G_{n}$ without
colliding with any vertical walls (Fig.~\ref{fig4}). Sometimes we
describe this as: the \o\ reaches \emph{directly} the aperture
$G_{n}$.

\medskip\medskip


It is now easy to see that \cite{ddl}:

\begin{enumerate}

\item The backward evolution of $G_{n}$ can only \emph{split} once for
each of the $n-1$ singular vertices (Fig.~\ref{fig4}). This implies
that $E_{\vartheta}^{(n)}$ is the union of at most $n$ right-open
intervals.  We denote this by $n.i.(E_{\vartheta}^{(n)}) \le n$, where
$n.i.$ stands for ``number of intervals''.

\item $|E_{\vartheta}^{(n)}| = 2p_{n}$.

\item $E_{\vartheta}^{(n+1)} \subset E_{\vartheta}^{(n)}$.  In
particular, the family $\{ E_{\vartheta}^{(n)} \}_{n>0}$ can be
rearranged into sequences of nested right-open intervals, whose
lengths vanish as $n \to\infty$.

\item If $E_{\vartheta} := \bigcap_{n>0} E_{\vartheta}^{(n)}$ denotes
the subset of $L$ on which $\Pm_{\vartheta}$ is not defined, then
clearly $|E_{\vartheta}|=0$.

\item Each point of $E_{\vartheta}$ is the limit of an infinite
sequence of nested vanishing right-open intervals. Moreover, each
infinite sequence yields a point of $E_{\vartheta}$, unless the
intervals eventually share their right extremes.

\end{enumerate}

Orbits starting at points of $E_{\vartheta}$ will never collide
with any vertical side of $\Su(\vartheta)$ (or $\Bi$) and thus, as
$t\to +\infty$, will go to infinity, maintaining a positive constant
$X$-velocity.  We call them \emph{\eo s}.
\skippar

It turns out that if the ``infinite cusp'' of a step polygon
narrows down very quickly, several interesting facts can be shown:

\begin{theorem}
	If the heights $\{ p_{n} \}$ of an infinite step \bi\ $\Bi$
	verify $p_{n+1} \le \lambda p_{n}$, with 
	\begin{displaymath} 
		0 < \lambda < \lambda_{0} := \frac{\sqrt{6}-1}{5} 
		\simeq 0.290\ldots, 
	\end{displaymath} 
	then, for almost all directions $\vartheta \in ]0,\pi/2[$, 
	there exists a subsequence $\{ n_{j} \}$ such that 
	$n.i.( E_{\vartheta}^{(n_{j})} ) = 1$.
	\label{thm-unique}
\end{theorem}

\begin{corollary}
	For almost all directions there is exactly one \eo.
	\label{cor-unique}
\end{corollary}

Let us name $\eta_{\vartheta}$ the (unique) \eo.  Also, with a certain
lack of originality, let us call \emph{typical} a direction
$\vartheta$ for which the above statements hold.

Theorem \ref{thm-unique} and Corollary \ref{cor-unique} are the
analogues of Corollaries 3 and 4 of \cite{ddl}, although their proofs,
which are to be found in Section \ref{escape}, are quite different. In
our previous article we exploited thoroughly the arithmetic properties
of the \emph{exponential \bi} $p_n=2^{-n}$, whereas here, to achieve a
certain generality, we use rather rough \me-theoretic estimates on
some sets of directions. This is why we are pretty confident that the
same results can indeed be proven for a much wider class of \bi s (see
the proof of Lemma \ref{lemma-unique} and in particular estimate
(\ref{unique-150})).

The next two results are also formulated for a \bi\ as in the
statement of Theorem \ref{thm-unique}.

\begin{lemma}
	For a.a. $\vartheta$, $\eta_{\vartheta}$ does not intersect
	any vertex.
	\label{lemma-past}
\end{lemma}

The above lemma, too, is proven in Section \ref{escape}. Not only does
it describe the behavior of the \eo, but more importantly, in
conjunction with Theorem \ref{thm-unique}, yields the following
assertions---which follow from the above in the same way as the entire
Section 3 of \cite{ddl} follows from Corollary 3 and Lemma 8.

\begin{proposition}
	Fixed a typical direction $\vartheta$,
	\begin{itemize}
		\item[(i)] $\eta_{\vartheta}$ is oscillating in the
		past.
		\item[(ii)] $\eta_{\vartheta}$ is contained in the 
		$\omega$-limit of every \o\ (except for itself).
		\item[(iii)] $\eta_{\vartheta}$ is contained in the 
		$\alpha$-limit of every \o\ (except for the \o\ that
		escapes in the past).
		\item[(iv)] Excluding the exceptions mentioned above,
		every semi-\o\ stays close to $\eta_{\vartheta}$ for an
		arbitrary long time.
		\item[(v)] Every invariant continuous function on
		$\Su(\vartheta)$ is constant. 
		\item[(vi)] The flow is minimal \iff
		$\eta_{\vartheta}$ is dense.
		\item[(vii)] The closure of $\eta_{\vartheta} \cap L$, 
		which is the ``trace'' of the \eo\ on the usual 
		Poincar\'e section, is either the entire $L$ or a 
		Cantor set.
	\end{itemize}
	\label{prop-from-ddl}
\end{proposition}

\begin{remark}
	Proposition \ref{prop-from-ddl},(iii) fixes a minor mistake in
	Corollary 6 of \cite{ddl}: it is obvious that assertion (i)
	there does not work for the (unique) \o\ that escapes in the
	past.
\end{remark}

\subsection{Ergodic Properties} \label{results-erg}

We now turn to the ergodic properties of our step billiards, and we
start by presenting a slight enhancement of a previous result.

\begin{theorem}
	Fix $\alpha = \tan\vartheta \not\in \Q$.  For every positive,
	monotonically vanishing sequence $\{ \bar{p}_{n} \} \subset
	\Q$, and every integer $k$, there exists a decreasing sequence 
	$\{ p_{n} \} \subset \Q$, with
	\begin{eqnarray*}
		p_{n} = \bar{p}_{n}, && \mathrm{for}\ 0 \le n \le k; \\  
		0 < p_{n} \le \bar{p}_{n}, && \mathrm{for}\ n > k; \\
		\sum_{n} p_{n} < \infty. &&
	\end{eqnarray*}
	such that the \bi\ flow $\fat$ on $\Su^{\Bi}(\vartheta)$, for 
	$\Bi \simeq \{ p_{n} \}$, is \erg\ (hence almost all
	\o s are dense).  
	\label{thm-fast-decay}
\end{theorem}

\proof Exactly the same as \cite{ddl}, Theorem 1, except that the
induction starts at $n=k$.
\qed

Theorem \ref{thm-fast-decay} provides an example (many, as a matter of
fact) of a \bi\ that is \erg\ in one direction. One would like to get
more: a \bi\ \erg\ for a.a.~directions, like the rectangular table.
We adapt to the infinite step \bi s some of the techniques presented
in \cite{tr} to show that this is true ``in general'', in the sense of
\cite{kms}.

More precisely, let $\s$ be the space of all step polygons with unit
area. (This means that, in this section, we drop the unsubstantial
requirement $p_{0}=1$.) Since each step polygon $P$ is uniquely
determined by the sequence of the heights of its vertical sides
$\{p_{n}\}_{n\in\N}$, we apply to $\s$ the metric of the space
$l^{1}$: given $P \simeq \{p_{n}\}$, $Q \simeq \{q_{j}\}$, let
\begin{displaymath}
	d(P,Q) := \sum_{n=0}^{\infty} |p_{n}-q_{n}|.
\end{displaymath}
The metric space $(\s,d)$ is complete and separable. Note that
$d(P,Q)=A(P\triangle Q)$.

We call a set of a topological space \emph{typical} if it contains a
dense $G_{\delta}$-set, \emph{meager} if it is the union of a
countable collection of nowhere dense sets. The following 
statement is the highlight of this section: 

\begin{theorem} \label{gigiTheorem} 
	A typical step polygon $P$ is ergodic on $\Su^{P} (\vartheta)$
	for a.e. $\vartheta \in ]0, \pi/2[$.
\end{theorem} 

\begin{corollary} \label{gigiCorollary}
	There are infinite step billiards $P$ ergodic on 
	$\Su^{P}(\vartheta)$ for a.e. $\vartheta \in ]0, \pi/2[$.
\end{corollary}

Both results are proven in Section \ref{ergodicity}. The main
differences between Theorem \ref{gigiTheorem} and Theorem 5.1 of
\cite{tr} are:

\begin{enumerate}
	\item He came first.

	\item The result in \cite{tr} is about a broad class of
	rational infinite polygons, while ours is restricted to the
	infinite step polygons.

	\item Even though they use the same ideas, Theorem
	\ref{gigiTheorem} cannot be derived by its analogue in
	\cite{tr}.

	\item His \bi s are bounded; ours can be unbounded.

	\item The metric that we use is very simple and corresponds to
	the intuitive idea of closeness.
\end{enumerate}

\subsection{Metric Entropy} \label{results-entropy}

Let $\Bi \simeq \{p_n\}$ be a given infinite step billiard. We denote by
${\cal V}=\{V_k\}$ the set of non-removable (singular) vertices,
defined at the beginning of Section \ref{notations}. One can prove the
following property of non-periodic orbits, which turns out to be very
useful in certain entropy estimates.

\begin{proposition} \label{limitpoint}
	The closure of the forward and backward semi-orbit of every
	non-periodic point intersects ${\cal V}\cup\{\infty\}$.
\end{proposition}

\begin{remark}
	This result can be considered the adapted version of
	Proposition 4.1 in \cite{tr}. Notice that in our case the
	billiard is always rational (or weakly rational according to
	\cite{tr}, Definition 4.3) and ${\cal V}$ is only a subset of
	all vertices. Also, it is immediate to check that for \bi s as
	in Section \ref{results-erg}, Proposition \ref{limitpoint} is
	contained in Proposition \ref{prop-from-ddl}.
\end{remark}

\proofof{Proposition \ref{limitpoint}} If a semi-orbit $\gamma$
(either forward or backward) of a non-periodic point is bounded, then
it is entirely contained in a finite step billiard $\Bin$, for $n$
large enough. According to \cite{gkt}, the closure of $\gamma$ must
contain a singular vertex of $\Bin$, which clearly is in ${\cal
V}$. If instead $\gamma$ is unbounded, then the set of its limiting
points contains $\{\infty\}$.
\qed

Take $n\ge 1$. In Fig.~\ref{fig2}, let $L_{n} := \{n\} \times
[p_n,p_{n-1}[$ and $L_{-n} := \{n\} \times [-p_{n-1},-p_n[$ be the two
copies of the $n$-th vertical side. We identify them with $d_{n} :=
[p_n,p_{n-1}[$ and $d_{-n} := [-p_{n-1},-p_n[$. The family of these
intervals partitions $I := [-1,0[ \cup ]0,1[$. Thus it makes sense to
define $f_\vartheta$, as the i.e.t. induced by $\fat$ on $I$.

For any $f_\vartheta$-invariant Borel probability measure $\nu$, we
set
\begin{displaymath}
	H_\nu (\Bi) := -\sum_{n=1}^{\infty}\,\nu(d_n)\,\log
	\nu(d_n).
\end{displaymath}
Call ${\tilde{\cal X}}_\vartheta := L \setminus \bigcap_{n=0}^{\infty}
\Pm_{\vartheta}^{-n} E_{\vartheta}$ the set of the points in $L$ whose
forward \o s keep returning there; and denote by ${\cal X}_\vartheta$
the corresponding set in $I$. More precisely,
\begin{displaymath}
	{\cal X}_\vartheta := g_\vartheta \left( {\tilde{\cal
	X}}_\vartheta \right),
\end{displaymath}
where $g_\vartheta \,:\, {\tilde{\cal X}}_\vartheta \longrightarrow I$
is given by $g_\vartheta(x)=\phi_{\vartheta,t_1}(x)$ and $t_1$ is the
first collision time at a vertical wall. For any $x\in {\tilde{\cal
X}}_\vartheta$, let
\begin{displaymath}
	{\tilde\pi}_{\vartheta}(x) := \{\ldots,L_{\omega_{-1}}, L,
	L_{\omega_0}, L, L_{\omega_1}, L, \ldots, L, L_{\omega_k},
	\ldots \},
\end{displaymath}
be the sequence of vertical sides that the \o\ of $x$ crosses (adopting
the convention that $L_{\omega_0}$ is the first vertical side
encountered in the past). Then we define the \emph{coding}
$\pi_\vartheta \,:\, {\tilde{\cal X}}_{\vartheta} \longrightarrow
(\Z^*)^{\Z}$ by
\begin{displaymath}
	\pi_\vartheta(x) := \{\ldots, \omega_{-1}, \omega_0, \omega_1,
	\ldots \}
\end{displaymath}
We equip $C_{\vartheta} := \pi_{\vartheta} ({\cal X}_{\vartheta})$
with the product topology and we denote by $\sigma$ the left
shift. The following diagram commutes:
\begin{displaymath}
\begin{array}{ccc}
	f_{\vartheta}: {\cal X}_\vartheta & \longrightarrow & 
	{\cal X}_\vartheta \\
 	\ \quad \uparrow & g_\vartheta & \uparrow  \\
	{\cal P}_\vartheta: {\tilde {\cal X}}_\vartheta & 
	\longrightarrow & {\tilde {\cal X}}_\vartheta \\
 	\ \quad \downarrow & \pi_\vartheta & \downarrow  \\
	\sigma : C_\vartheta & \longrightarrow & C_\vartheta
\end{array}
\end{displaymath}

\begin{proposition} \label{entropy}
	Let $\nu$ be any $f_\vartheta$-invariant, Borel, \erg\
	probability \me\ on ${\cal X}_\vartheta$. If $H_\nu (\Bi)
	<\infty$, then $h_\nu=0$.
\end{proposition}

\begin{remark}
	For instance, if $p_{n}$ decays exponentially or faster, as in
	the examples that we have recalled earlier, then $H_{\ell}
	(\Bi)$ is finite for the Lebesgue \me\ $\ell$.
\end{remark}

\proofof{Proposition \ref{entropy}} First of all, notice that the
assertion is obvious if $\nu$ is atomic. We claim that the partition
$\alpha := \{ \alpha_1, \alpha_2, \ldots \}$ given by
\begin{displaymath}
	\alpha_j := \{\omega\in C_\vartheta \st \omega_1\,=\,j \}
\end{displaymath}
is a one-side generator. This implies by standard results (e.g.,
\cite{r}, 11.3) that $h_{\lambda}=0$ for any $\sigma$-invariant Borel
probability \me\ $\lambda$ on $C_\vartheta$.  Now, $g_\vartheta$ is
bijective by construction, therefore the pull-back $g_{\vartheta *}
\nu$ is also non-atomic and \erg. Hence, there is a $Y \subseteq
{\tilde {\cal X}}_\vartheta$ such that $\pi_\vartheta|_{Y}$ is
injective and $g_{\vartheta *} \nu (Y) = 1$. We conclude that $h_\nu
(f_\vartheta) = h_{\pi_\vartheta^{*} g_{\vartheta *} \nu} (\sigma) =
0$.

It remains to prove the initial claim. This is done if we show that
the sequence of sides an orbit visits in the future determines the
sequence of sides visited in the past.  Clearly, if the orbit is
periodic, there is nothing to prove.  In the non-periodic case we can
use Proposition \ref{limitpoint} to conclude that any parallel strip
of orbits must eventually ``split'' at some vertex $V_k$ and this of
course implies that two distinct orbits cannot hit the same vertical
sides in the future.  
\qed

We now turn to the proofs of the other results.

\section{Escape orbits} \label{escape}

In this section we denote directions by $\alpha = \tan\vartheta \in
]0,+\infty[$, using the Lebesgue \me\ there.  All surfaces $S(\alpha)$
will be identified with the set $S$ of Fig.~\ref{fig2} on which we use
the coordinates $(X,Y)$. For instance, $\gamma_k(\alpha)$ will denote
the \o\ on $S(\alpha)$ starting at point the $(k,-p_k) \simeq
V_k$---this is the same as the \o\ on $S$ starting at $(k,-p_k)$ with
slope $\alpha$.  
\skippar

\proofof{Theorem \ref{thm-unique}} Let $k<m$ be two natural
numbers. We introduce the following sets:
\begin{eqnarray}
	A_{k,m} &:=& \{ \a \st \gamma_{k}(\a) \mbox{ reaches directly
	} G_{m} \};  \label{unique-2}  \\
	B_{m} &:=& \bigcup_{k=1}^{m-1} A_{k,m}.  \label{unique-4}
\end{eqnarray}
Therefore,
\begin{displaymath}
	B_{m}^{c} = \{ \a \st \mbox{no } \gamma_{k}(\a), \mbox{
	with }1 \le k \le m-1, \mbox{ reaches directly } G_{m} \}.
\end{displaymath}
Also, let us define
\begin{eqnarray*}
	C &:=& \bigcap_{n\in\N} \bigcup_{m\ge n} B_{m}^{c} =  \\
	&=& \{ \a \st \exists \: \{ n_{j} \}\ s.t.\ G_{n_{j}}
	\mbox{ is not reached by any } \gamma_{k}(\a), k \le n_{j}
	\}. 
\end{eqnarray*}
When such a subsequence exists, the backward evolution of $G_{n_{j}}$
does not split at any of the vertices $V_{k},\, 1\le k\le n_{j}$.
Hence $n.i.( E_{\a}^{(n_{j})} ) = 1$. Therefore establishing the
theorem amounts to proving that $|C^{c}| = 0$. This is implied by the
following:
\begin{equation}
	\forall\in\N, \quad \left| \bigcap_{m\ge n} B_{m} \right| = 0.
	\label{unique-10}
\end{equation}
In order to obtain (\ref{unique-10}), we introduce some notation, and a
lemma. Given two sets $A,I$, with $I$ bounded, denote $|A|_{I} := |A\cap
I|/|I|$.

\begin{lemma}
	Assumptions and notations as in Theorem \ref{thm-unique}. For
 	every (bounded) interval $I \subset \R^{+}$, there exists a
 	$\delta \in ]0,1[$ such that $\limsup_{m\to\infty} |B_{m}|_{I}
 	\le \delta$.  
	\label{lemma-unique}
\end{lemma}

The proof of this lemma will be given in the sequel. Now, proceeding by
contradiction, let us suppose that $\left| \bigcap_{m\ge n} B_{m}
\right| \ne 0$, for some $n$. By the Lebesgue's Density Theorem, almost
all points of that set are points of density. Pick one: this means that
there exists an interval $I$, around that point, such that
\begin{displaymath}
	\left| \bigcap_{m\ge n} B_{m} \right|_{I} \ge \sigma > \delta.
\end{displaymath}
Hence, $\forall m\ge n,\ |B_{m}|_{I} \ge \sigma$, which contradicts the
lemma. This proves (\ref{unique-10}) and Theorem \ref{thm-unique}.
\qed

\begin{figure}[ht]
    \hspace{2cm}
    \includegraphics[width=4.0in]{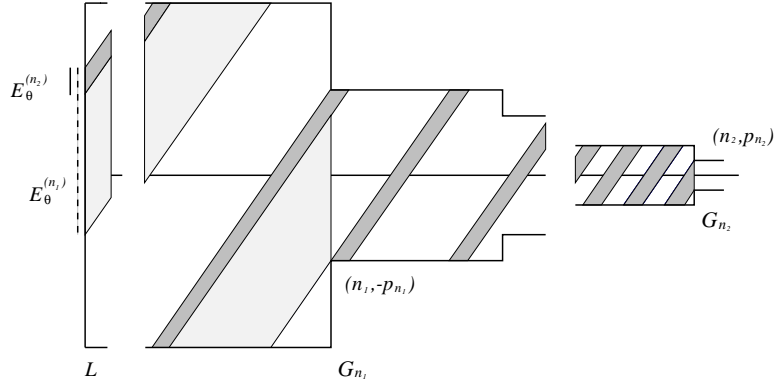}
    \caption{\small The fact that $E_{\vartheta}^{(n_{1})}$ and
 $E_{\vartheta}^{(n_{2})}$ have upper (equivalently right) extremes
in common implies the existence of a generalized diagonal.}
    \protect\label{fig5}
\end{figure}

\proofof{Corollary \ref{cor-unique}} The previous theorem implies
that, for a.a. $\a$'s, $\# E_{\a} =$ 0 or 1. As already mentioned, the
former case (no \eo s), occurs \iff the intervals $E_{\a}^{(n_{j})}$
share their right extremes, for $j$ large. This implies the existence
of generalized diagonals (see Fig.~\ref{fig5}). Excluding those cases
only amounts to removing a null-\me\ set of directions.
\qed

\medskip\medskip
\medskip\medskip

\begin{figure}[ht]
    \hspace{2cm}
    \includegraphics[width=4.0in]{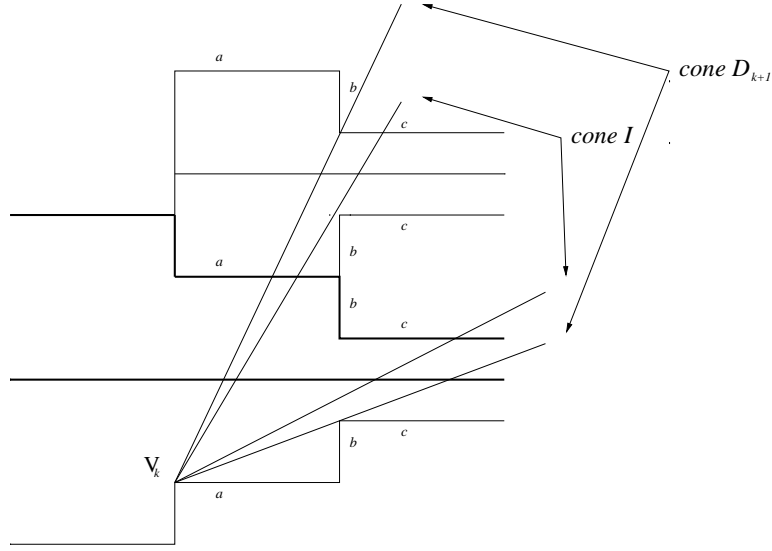}
   \caption{\small Unfolding of the billiard. Trajectories 
 departing from a given singular vertex $V_{k}$ are drawn as
 straight lines on the plane. Every time one of these hits a side
of the billiard, a new copy of the billiard, reflected around that
 side, is drawn. Cones $I$ and $D_{k+1}$ are used in the proof of
Lemma \ref{lemma-unique}.}
    \protect\label{figm1}
\end{figure}

\proofof{Lemma \ref{lemma-unique}} In this proof we will heavily
use the technique of \bi-unfolding; that is, in order to draw an
\o\ as a straight line in the plane, we reflect the \bi\ around
one of its sides every time the \o\ hits it, as shown in
Fig.~\ref{figm1}. 

\medskip\medskip


Fix $k$, and view $I$ as a conical beam of \tr ies departing from
$V_{k}$ (Fig.~\ref{figm1}): this makes sense since these \tr ies are
in one-to-one correspondence with their slopes. The goal is to exploit
the geometry of the unfolded \bi\ to set up a recursive argument that
will yield exponential bounds for $|A_{k,m}|_{I}$ (in $m$).

\begin{figure}[ht]
    \hspace{2cm}
    \includegraphics[width=4.0in]{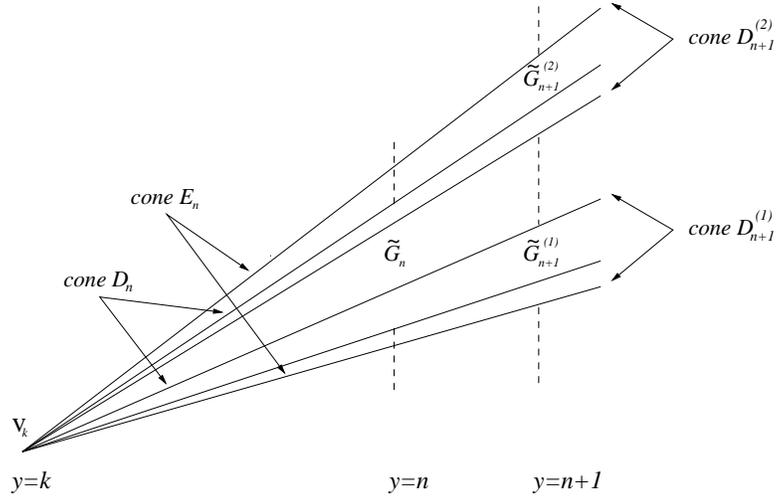}
  \caption{\small Proof of Lemma \ref{lemma-unique}: 
construction of the cones $D_{n}$ and $E_{n}$ in a roughly
 sketched unfolded-\bi\ plane.}
    \protect\label{figm2}
\end{figure}

Here our recursive argument starts. In the unfolded-\bi\ plane,
sketched in Fig.~\ref{figm2}, let $l_{k}(\a)$ be the straight line of
slope $\a$ passing through $V_{k}$. Take an $n>k$ and consider one
copy of $G_{n}$, indicated as an ``opening'' in Fig.~\ref{figm2}:
$\tilde{G}_{n} := \{n\} \times [r-p_{n}, r+p_{n}[$, for some $r\in
\R^{+}$. The straight lines (departing from $V_{k}$) that cross
$\tilde{G}_{n}$ encounter a $2p_{n}$-periodic array of copies of
$G_{n+1}$. We call them $\tilde{G}_{n+1}^{(j)} := \{n+1\} \times
[r-p_{n+1}+j \cdot 2p_{n}, r+p_{n+1}+j \cdot 2p_{n}[$; $j$ assumes a
finite number (say $\ell$) of integer values. Define the interval
\begin{equation}
	D_{n} := \{ \a \st l_{k}(\a) \cap \tilde{G}_{n} \ne
	\emptyset \}.
	\label{unique-30}
\end{equation}

\medskip\medskip


\begin{remark}
	It might be convenient to think of the elements of $D_{n}$ as
	lines sharing the common point $V_{k}$: specifically the lines
	in $D_{n}$ are those that cross $\tilde{G}_{n}$. So it makes
	sense to refer to $D_{n}$ (and similar sets) as a
	cone. (Incidentally, $|D_{n}| = 2p_{n}/(n-k)$.)  On the other
	hand, the one-to-one correspondence between $\a \in \R^{+}$
	and $\gamma_{k}(\a)$, on one side, and $l_{k}(\a)$, on the
	other, should not lead one to think that the latter is the
	``lifting'' of the former by means of the unfolding
	procedure. This is only the case when $\gamma_{k}(\a)$ reaches
	$G_{n}$. As a matter of fact, the inclusion $A_{k,n} \cap
	D_{n} \subseteq D_{n}$ is expected to be strict for general
	choices of $D_{n}$, $n>k+1$.
\end{remark}

The cone $D_{n}$ cuts a segment on the vertical line $y=n+1$. This
segment includes some $\tilde{G}_{n+1}^{(j)}$ and only intersects some
other $\tilde{G}_{n+1}^{(i)}$ (at most two, of course).  Now expand
$D_{n}$ in such a way that the segment includes $\ell$ ``full'' copies
of $G_{n+1}^{(j)}$; the resulting cone will be denoted by
$E_{n}$. Fig.~\ref{figm2} shows that in this operation we might have
to attach, on top and on bottom of $D_{n}$ two cones of \me\ up to
$2p_{n+1}/(n+1-k)$.  Therefore:
\begin{equation}
	\frac{|E_{n}|}{|D_{n}|} \le 1 +
	\frac{4p_{n+1}}{(n+1-k)|D_{n}|} \le \frac{2p_{n+1} + p_{n}}
	{p_{n}}.
	\label{unique-40}
\end{equation}
We further define the set
\begin{equation}
	D_{n+1} := \{ \a \in E_{n} \st l_{k}(\a) \cap
	\tilde{G}_{n+1}^{(j)} \ne \emptyset, \mbox{ for some } j \} =
	\bigcup_{i=1}^{\ell} D_{n+1}^{(i)},
	\label{unique-50}
\end{equation}
where the $D_{n+1}^{(i)}$ are cones of measure $2p_{n+1}/(n+1-k)$. One
has:
\begin{equation}
	\frac{|D_{n+1}|}{|E_{n}|} \le \frac{2\cdot 2p_{n+1}} {2\cdot
	2p_{n+1} + 2(p_{n}-p_{n+1})} = \frac{2p_{n+1}} {p_{n+1}+p_{n}}.
	\label{unique-60}
\end{equation}
In fact it is not hard to realize that the l.h.s.~of (\ref{unique-60})
is largest when $\ell=2$. In that case, $E_{n} \setminus D_{n+1}$ is a
interval of \me\ $2(p_{n}-p_{n+1})/(n+1-k)$, whence the second term of
(\ref{unique-60}). Combining (\ref{unique-40}) and (\ref{unique-60})
we obtain
\begin{equation}
	\frac{|D_{n+1}|}{|D_{n}|} \le \frac{2p_{n+1} (2p_{n+1} + p_{n})}
	{p_{n} (p_{n+1} + p_{n})} =: \beta_{n}.
	\label{unique-70}
\end{equation}
At this point we notice that each $D_{n+1}^{(i)}$ as introduced in
(\ref{unique-50}) is again a set of the type (\ref{unique-30}), with
$n+1$ replacing $n$.  Hence estimate (\ref{unique-70}) holds and
$|D_{n+2}^{(i)}| \le \beta_{n+1} |D_{n+1}^{(i)}|$ with $D_{n+2}^{(i)}$
suitably defined as in the above construction.  Call $D_{n+2} =
\cup_{i=1}^{\ell} D_{n+2}^{(i)}$: this set is also a union of cones of
equal size.  One has
\begin{displaymath}
	\frac{|D_{n+2}|}{|D_{n+1}|} \le \frac{\sum_{i=1}^{\ell}
	|D_{n+2}^{(i)}|} {\ell \, |D_{n+1}^{(i)}|} \le \beta_{n+1},
\end{displaymath}
and the trick can continue.

We are now ready to implement the recursive argument: assume that some
of the \o s of the cone $I$ (based in $V_{k}$) cross $G_{k+1}$, that
is, $A_{k,k+1} \cap I \ne \emptyset$ (if not, everything becomes
trivial as we will see later). In the unfolded-\bi\ plane, enlarge
$A_{k,k+1} \cap I$ until it fits the minimal number of copies of
$G_{k+1}$, as in Fig.~\ref{figm1}; call this new set $D_{k+1}$.  This
is by definition a finite union of intervals of the type
(\ref{unique-30}), of fixed size). Hence inequality (\ref{unique-70})
applies and the definition/estimate algorithm can be carried on until
we define, say, $D_{m}$. The only thing we need to know about this set
is that $(A_{k,m} \cap I) \subseteq D_{m}$, which should be clear by
construction (see also the previous remark). This fact and the
repeated use of (\ref{unique-70}), yield
\begin{equation}
	|A_{k,m} \cap I| \le |D_{k+1}| \prod_{i=k+1}^{m-1} \beta_{i} .
	\label{unique-90}
\end{equation}

Let us consider our specific case: $p_{n+1} \le \lambda p_{n}$. From
definition (\ref{unique-70}), one verifies that:
\begin{equation}
	\beta_{n} = \frac{2 (2 (p_{n+1}/p_{n}) + 1)} {((p_{n}/p_{n+1})
	+ 1)} \le \frac{2(2\lambda + 1)}{\lambda^{-1} + 1} =: \beta.
	\label{unique-100}
\end{equation}
It is going to be crucial later that $\beta$ be less than 1. For
$\lambda$ positive, this amounts to $4\lambda^2 +\lambda -1 <0$, which
is easily solved by
\begin{equation}
	0 < \lambda < \lambda_{1} := \frac{\sqrt{17}-1}{8} \simeq
	0.390\ldots,
	\label{unique-110}
\end{equation}
Going back to the definition of $D_{k+1}$, and to (\ref{unique-90}),
we see that it is possible to give an estimate of the \me\ of
$D_{k+1}$ in $I$, for large $k$. In fact, when $|I|$ is much bigger
than $2p_{k}$, then it is clear (from Fig.~\ref{figm1}, say) that $I$
includes very many cones of size $2p_{k+1}$, placed on a
$2p_{k}$-periodic array. As $k$ grows, the density of these cones in
$I$ can be made arbitrarily close to $p_{k+1}/p_{k} \le \lambda$. The
precise statement then is: given any $\eps>0$, there exists a $q =
q(\eps) \in\N$ such that
\begin{displaymath}
	\forall k\ge q, \quad |D_{k+1}| < (\lambda + \eps) \, |I|.
\end{displaymath}
Plugging this into (\ref{unique-90}), we obtain
\begin{equation}
	\forall m>k\ge q \quad |A_{k,m}|_{I} \le (\lambda + \eps) \,
	\beta^{m-k-1},
	\label{unique-130}
\end{equation}
having used (\ref{unique-100}) as well. For the other values of $k$,
we have no control over $C_{k} := |D_{k+1}|/|I|$ and we just write
\begin{equation}
	\forall m,q > k \quad |A_{k,m}|_{I} \le C_{k}\, \beta^{m-k-1}.
	\label{unique-140}
\end{equation}
In the case $A_{k,k+1} \cap I = \emptyset$, which we did not consider
before, (\ref{unique-130})-(\ref{unique-140}) are trivial consequences
of the fact that $A_{k,m} \cap I = \emptyset$, for all $m>k$.

We move on to the final estimation. Take $m>q$: from definition
(\ref{unique-4}) we have, using (\ref{unique-130}) and
(\ref{unique-140}),
\begin{eqnarray}
	|B_{m}|_{I} &\le& \sum_{k=1}^{q-1} |A_{k,m}|_{I} +
	\sum_{k=q}^{m-1} |A_{k,m}|_{I} \le  \nonumber  \\
	&\le& \sum_{k=1}^{q-1} C_{k}\, \beta^{m-k-1} + \sum_{k=q}^{m-1}
	(\lambda + \eps) \, \beta^{m-k-1} \le  \label{unique-150}  \\
	&\le& o(1) + \frac{\lambda + \eps}{1-\beta},  \nonumber
\end{eqnarray}
as $m \to \infty$. In the last inequality we have used twice the fact that
$\beta<1$. We impose the condition
\begin{equation}
	\frac{\lambda}{1-\beta} = \frac{\lambda (\lambda + 1)}
	{-4\lambda^{2} -\lambda +1} < 1.
	\label{unique-160}
\end{equation}
For $\lambda$ as in (\ref{unique-110}), the denominator is positive,
hence (\ref{unique-160}) can be rewritten as $5\lambda^{2} +2\lambda -1
<0$, whose solutions are
\begin{equation}
	0 < \lambda < \frac{\sqrt{6}-1}{5} = \lambda_{0} < \lambda_{1},
\end{equation}
as in the statement of the lemma. For these values of $\lambda$, by
virtue of (\ref{unique-110}), $1-\beta$ keeps away from 0. Therefore
(\ref{unique-160}) implies that, for $\eps$ small enough, the last
term in (\ref{unique-150}) can be taken less than a certain $\delta <
1$, whence the proof of Lemma \ref{lemma-unique}.
\qed

\proofof{Lemma \ref{lemma-past}} Consider a vertex $V$ of $\Bi_{\a}$
and let $\gamma_{V}(\a)$ be its forward semi-\o. For a fixed finite
sequence $S:= (S_{1}, S_{2}, \ldots S_{\ell})$ of sides, we define
\begin{equation}
	A_{V,m}(S) := \{ \a \st \gamma_{V}(\a) \mbox{ hits } S_{1},
	\ldots, S_{\ell} \mbox{ and then reaches directly
	} G_{m} \}.
	\label{past-10}
\end{equation}
This means that these \tr ies do not hit any vertical wall after
leaving $S_{\ell}$ and before reaching $G_{m}$. Notice the
similarities with definition (\ref{unique-2}). As a matter of fact, if
$V=V_{k}$ for some $k$, then $A_{V_{k},m}(\emptyset) =
A_{k,m}$. However, for most sequences $S$, (\ref{past-10}) defines the
empty set. For example, $S$ can be \emph{incompatible} in the sense
that no \o\ can go from $S_{i}$ to $S_{i+1}$ without crossing other
sides in the meantime. But, even for compatible sequences, if $G_{m}$
does not lie to the right of $S_{\ell}$, obviously $A_{V,m}(S) =
\emptyset$. To avoid this latter case, we fix $m_{o} = m_{o}(V,S)$
bigger than the largest $Y$-coordinate in $S_{\ell}$. For $m \ge
m_{o}$, $A_{V,m+1}(S) \subset A_{V,m}(S)$. Let us then define
\begin{eqnarray}
	B_{V}(S) &:=& \{ \a \st \gamma_{V}(\a) \mbox{ hits } S_{1},
	\ldots, S_{\ell} \mbox{ and then escapes to } \infty \} =
	\nonumber  \\
	&:=& \bigcap_{m=m_{o}}^{\infty} A_{V,m}(S),  \label{past-20}
\end{eqnarray}
Working in the unfolded-\bi\ plane and identifying directions with \o
s, it is not hard to realize that $A_{V,m_{o}}(S)$ is made up of a
finite number of intervals/cones, each of which reaches a copy of
$G_{m_{o}}$ after hitting a certain sequence of sides $(S_{1}, \ldots
S_{\ell}, S_{\ell+1}, \ldots S_{n})$ (the first $\ell$ sides are
common to all cones and the others can only be horizontal).  It is
possible that some of these beams of \tr ies intersect the
corresponding copy of $G_{m_{o}}$ only in a proper sub-segment. Let us
fix this situation by enlarging any such beam until it covers the
whole segment. We call $D \supseteq A_{V,m_{o}}(S)$ the union of these
new cones.

Proceeding exactly as in the proof of Lemma \ref{lemma-unique} (see in
particular (\ref{unique-90}) and (\ref{unique-140})) we get, for $m >
m_{o}$,
\begin{displaymath}
	|A_{V,m}(S)| \le |D| \, \beta^{m-m_{o}},
\end{displaymath}
with $\beta < 1$. Hence, for $m \to \infty$, $|A_{V,m}(S)| \to 0$. By
(\ref{past-20}), $|B_{V}(S)| = 0$, and the set
\begin{displaymath}
	\bigcup_{V \atop \mathrm{vertex}} \bigcup_{S\ \mathrm{finite} \atop
	\mathrm{sequence}} B_{V}(S)
\end{displaymath}
has \me\ zero. This is the set of directions as in the statement of
the Lemma \ref{lemma-unique}, which we have now proved.
\qed

\section{Generic ergodicity} \label{ergodicity}

We begin this section by giving some definitions and a technical
lemma. In $\R^{2}$, let $\rho$ be the Euclidean metric and $A$ the
area. Also, recall the definition of $\s$ from Section
\ref{results-erg}; later on we will need to use $\s_{0}$, the
collection of all finite $\Bi \in \s$ with rational heights.

\begin{definition}
Given $P,Q \in \s$, $\vartheta \in ]0, \pi/2[$, $x \in
\Su^{P}(\vartheta) \cap \Su^{Q}(\vartheta)$ and
$\eps > 0$, let
\begin{eqnarray}
	&& I(P,Q,x,\eps) := \{t \in [0,1/\eps] \st
	\rho(\phi_{t}^{P}(x),\phi_{t}^{Q}(x)) > \eps \}, \label{I} \\
	&& G(P,Q,\vartheta,\eps) := \{ x \in \Su^{P}(\vartheta) \cap
	\Su^{Q}(\vartheta) \st |I(P,Q,x,\eps)| < \eps \}, \label{G} \\
	&& E(P,Q,\eps) := \left\{ \vartheta \in ]0, \pi/2[ \st 
	A(G(P,Q,\vartheta,\eps)) \ge A(\Su^{P,Q}(\vartheta)) -
	\eps \right\}. \label{E}
\end{eqnarray}
\end{definition}

\begin{lemma} \label{approx}
	For any $P \in \s$ and $\eps > 0$, we have
	\begin{displaymath}
		\lim_{Q \rightarrow P} |E(P,Q,\eps)|=1.
	\end{displaymath}
\end{lemma} 

\proofof{Lemma \ref{approx}} It is enough to prove the statement for
any sequence converging to $P$. Let $\{P_{n}\}$ be such a
sequence. Fix $\vartheta \in ]0, \pi/2[$ and $\eps >0$. Only countably
many \o s of $\Su ^{P}(\vartheta )$ contain a singular vertex $V$.
Let $x\in \Su^{P}$ be a point that does not belong to any of these \o
s. In a finite interval of time, its \tr y can get close only to a
finite number of singular vertices.  Therefore $\delta_{\eps}$, the
distance between the $\cup_{0\leq t\leq 1/\eps} \, \fat^{P}(x)$ and
$\mathcal{V}$, is positive. Let $N_{\eps}$ be the number of collisions
of $\cup_{0\leq t\leq 1/\eps} \, \fat^{P}(x)$ with the horizontal
sides of $\Su^{P} (\vartheta)$.

The two surfaces $\Su^{P}(\vartheta)$ and $\Su^{P_{n}}(\vartheta)$
overlap (as subsets of $\R^{2}$). If $x\in \Su^{P_{n}}$ as well, then
the set $\cup_{0\leq t\leq 1/\eps} \, \fat^{P_{n}}(x)$ represents
a finite piece of the trajectory of $x$ in the surface $\Su^{P_{n}}
(\vartheta)$.  We want to estimate $\rho (\phi^{P}_{t}(x),
\phi^{P_{n}}_{t}(x))$ for $0\leq t\leq 1/\eps$.  Since there is a
natural correspondence between the sides of the two surfaces, when we
say that the two trajectories hit the same sequence of sides, we mean
corresponding sides.  Notice that $\rho (\phi^{P}_{t}(x),
\phi^{P_{n}}_{t} (x))$ stays constant when neither \o\ crosses any
sides. Also, as long as $\phi^{P}_{t}(x)$ and $\phi^{P_{n}}_{t}(x)$
hit the same sequence of sides, every time that there is collision at
a horizontal side, the distance $\rho (\phi^{P}_{t}(x),
\phi^{P_{n}}_{t}(x))$ increases by a term $h \leq
2d(P,P_{n})$. Hence the maximum distance between the two \tr ies
in the interval $t \in [0, 1/\eps]$ is $\le 2 d(P,P_{n}) N_{\eps}$.
Therefore, by definition of $\delta_{\eps}$, if $d(P,P_{n}) <
\delta_{\eps} / (2N_{\eps})$, the \tr ies encounter the same sequence
of sides for $0\leq t\leq 1/\eps$.

Since $d(P,P_{n})\rightarrow 0$ as $n\rightarrow \infty $, we can find
an $m(x,\eps)>0$ such that the previous inequality is satisfied for
all $n>m(x,\eps)$. It is clear now that, as $n$ grows larger, the
distance between the two trajectories decreases. We conclude that, for
points $x\in \Su^{P}(\vartheta)$ with non-singular positive
semi-trajectory,
\begin{displaymath}
	\lim_{n\to\infty} \, \max_{0\leq t\leq 1/\eps} \rho
	(\phi^{P}_{t}(x), \phi^{P_{n}}_{t}(x)) = 0.
\end{displaymath}
As a consequence we have that a.e. $x\in \Su^{P} (\vartheta)$ belongs
to $G(P,P_{n},\vartheta,\eps)$ if $n$, which depends on $x$, is
sufficiently large. Therefore $A( G(P, P_{n}, \vartheta, \eps) ) \to
A(\Su^{P} (\vartheta))$ as $n\to \infty $. Being this true for all
$\vartheta \in ]0, \pi/2[$, we finally obtain $|E(P,P_{n},\eps)|\to 1$
as $n\to \infty$, for any $\eps >0$. 
\qed

We are now in position to attack the main proof of this section:
\skippar 

\proofof{Theorem \ref{gigiTheorem}} Choose $\eps_{n}>0$ such that
$\lim_{n\to \infty} \eps_{n}=0$. Let $\{ f_{i} \}_{i\in \N}$ be a
countable collection of continuous functions with compact support and
let it be dense in $L^{1}(\R^{2})$. For each step polygon $P$,
$f_{i}^{P}$ denotes the restriction of $f_{i}$ to
$\Su^{P}(\vartheta)$, corrected to obey the identifications on
$\Su^{P}(\vartheta)$. These corrections occur on a set of zero
Lebesgue measure in $\R^{2}$, therefore $\{f^{P}_{i}\}_{i\in \N }$ is
dense in $L^{1}(\Su^{P}(\vartheta))$ for each $\vartheta \in ]0,
\pi/2[$.

Given $P\in \s$, $x\in \Su^{P} (\vartheta)$ and $i\in \N$, let us introduce
\begin{displaymath}
	B_{T}^{P} (\vartheta,i,x) := \left| \frac{1}{T} \int_{0}^{T}f_{i}^{P}
	\circ \phi^{P}_{t}(x) dt - \frac{1} {A(\Su^{P}(\vartheta))}
	\int_{\Su^{P}(\vartheta)} f_{i}^{P} dA \right|
\end{displaymath}
and
\begin{displaymath}
	C_{T}^{P} (\vartheta,n) := \left\{ x\in \Su^{P}(\vartheta) \st
	B_{T}^{P} (\vartheta,i,x) \leq \eps_{n},\: i=1,\ldots, n
	\right\}.
\end{displaymath}
If $P\in \s_{0}$, the billiard flow $\fat^{P}$ is uniquely ergodic for
all irrational $\vartheta$ (i.e., $\tan \vartheta \not\in \Q$).  Thus
for every $n>0$ and every irrational $\vartheta \in ]0, \pi/2[$, we
have that $\lim_{T\to +\infty} A(C_{T}^{P} (\vartheta,n)) = A(\Su^{P}
(\vartheta))$. Let
\begin{displaymath}
	D_{T}^{P}(n) := \left\{ \vartheta \in ]0, \pi/2[ \st
	A(C^{P}_{T} (\vartheta,n)) \geq A(\Su^{P} (\vartheta)) -
	\eps_{n} \right\}.
\end{displaymath}
Then we can choose a $T_{n}(P)\geq n$ such that $|D_{T}^{P}(n)| > 1 -
\eps_{n}$ for any $T \geq T_{n}(P)$. Since $f_{1}, f_{2}, \ldots,
f_{n}$ are uniformly continuous on $\R^{2}$, there is an $r_{n}>0$ for
which $|f_{i}(x)-f_{i}(y)| \leq \eps_{n}$ whenever $\rho(x,y) \leq
r_{n}$ and $i=1, \ldots, n$. Let
\begin{displaymath}
	\delta_{n}(P) := \min \left\{ \frac{1}{T_{n}(P)}, \eps_{n},
	\frac{\eps_{n}} {\max_{1\leq i\leq n} \| f_{i} \|_{\infty} },
	r_{n} \right \}
\end{displaymath}
and $\tau_{n} := 1/\delta_{n}$. According to Lemma \ref{approx}, there
exists $0 < \sigma_{n}(P) \leq \delta_{n}(P)$ such that if $Q\in
U_{\sigma_{n}}(P) := \{R\in \s \st d(P,R)\leq \sigma_{n} \}$, then
$|E(P,Q,\delta_{n})| > 1 - \delta_{n} \geq 1 - \eps_{n}$.

Let $E_{n} := E(P,Q,\delta_{n})$ and $I_{n}(x)$, $G_{n}(\vartheta)$ be
the sets (\ref{I}), (\ref{G}) used in the definition of
$E(P,Q,\delta_{n})$. These sets depend on $P$.  For $\vartheta \in
E_{n}$ and $x\in G_{n}(\vartheta)$, $A(G_{n}(\vartheta)) \geq
A(\Su^{P,Q} (\vartheta)) - \delta_{n} \geq A(\Su^{P,Q} (\vartheta)) -
\eps_{n}$, $|I_{n}(x)| < \delta_{n}$ and $\rho (\phi^{P}_{t}(x),
\phi^{Q}_{t}(x)) \leq \delta_{n} \leq r_{n}$ for $t\in
[0,\tau_{n}]\setminus I_{n}(x)$. Notice that $T_{n} \leq \tau_{n}$.
For $i=1, \ldots, n$, we have:
 
\begin{eqnarray}
	&& \left| \frac{1} {A(\Su^{P})} \int_{\Su^{P}} f_{i}^{P} dA -
	\frac{1}{A(\Su^{Q})} \int_{\Su^{Q}} f_{i}^{Q} dA \right| 
	= \nonumber \\
	&& \qquad = \frac{1} {A(\Su^{P})} \left| \int_{\Su^{P}}
	f_{i}^{P} dA - \int_{\Su^{Q}} f_{i}^{Q} dA \right| \le 
	\label{espect} \\
	&& \qquad \le \frac{1}{4} \int_{\Su^{P} \triangle \Su^{Q}} 
	\left| f_{i} \right| dA \leq \frac{1}{4} \| f_{i} \|_{\infty}
	A(\Su^{P} \triangle \Su^{Q}) \le \nonumber \\ 
	&& \qquad \le \sigma_{n} \| {f_{i}} \|_{\infty} \leq
	\delta_{n} \| {f_{i}} \|_{\infty} \leq \eps_{n}. \nonumber
\end{eqnarray}
Let $\vartheta \in D_{\tau_{n}}^{P}(n)\cap E_{n}$ and $x\in
C^{P}_{\tau_{n}}(\vartheta,n)\cap G_{n}(\vartheta)$. Then
\begin{eqnarray*}
	&& \left| \frac{1}{\tau_{n}} \int^{\tau_{n}}_{0} f_{i}^{Q}
	\circ \phi_{t}^{Q}(x) dt - \frac{1}{A(\Su^{Q})} \int_{\Su^{Q}}
	f_{i}^{Q} dA \right| \\ 
	&& \qquad \le \frac{1}{\tau_{n}} \int^{\tau_{n}}_{0} \left| 
	f_{i}^{Q} \circ \phi^{Q}_{t}(x) - f_{i}^{P} \circ 
	\phi^{P}_{t}(x) \right| dt + \\
	&& \qquad + \left| \frac{1}{\tau_{n}} \int^{\tau_{n}}_{0} f_{i}^{P}
	\circ \phi^{P}_{t}(x) dt - \frac{1}{A(\Su^{P})} \int_{\Su^{P}}
	f_{i}^{P} dA \right| + \\
	&& \qquad + \left| \frac{1}{A(\Su^{P})} \int_{\Su^{P}} f_{i}^{P} dA -
	\frac{1}{A(\Su^{Q})} \int_{\Su^{Q}} f_{i}^{Q} dA \right| \\
	&& \qquad =: I+II+III.
\end{eqnarray*}
We have $I \leq 2\, \delta^{2}_{n} \| f_{i} \|_{\infty} + \eps_{n} \leq
3\eps_{n}$---say---for $n$ large enough. Moreover, $x\in
C^{P}_{\tau_{n}}(\vartheta,n)$ implies $II \leq \eps_{n}$. Finally,
$III \leq \eps_{n}$ by (\ref{espect}). We conclude that $I+II+III \leq
5\eps_{n}$ for $\vartheta \in D_{\tau_{n}}^{P}(n) \cap E_{n}$, $x \in
C^{P}_{\tau_{n}} (\vartheta,n) \cap G_{n}(\vartheta)$ and $i=1,
\ldots,n$. By definition of $D_{\tau_{n}}^{P}(n)$ and $E_{n}$, we
have:
\begin{equation} \label{directions}
	|D_{\tau_{n}}^{P}(n) \cap E_{n}| > 1 - 2\eps_{n}.
\end{equation}
From $A(C^{P}_{\tau_{n}} (\vartheta,n)) \geq A(\Su^{P}(\vartheta)) -
\eps_{n}$ and $A(G_{n}(\vartheta)) \geq A(\Su^{P}(\vartheta)) -
\eps_{n}$, it follows that
\begin{equation} \label{initcond}
	A(C_{\tau_{n}}^{P} (\vartheta,n) \cap G_{n}(\vartheta)) \geq
	A(\Su^{P} (\vartheta)) - 2\eps_{n} = A(\Su^{Q}(\vartheta)) -
	2\eps_{n}.
\end{equation}

Let $\{P_{j}\}_{j\in \N}$ be an enumeration of $\s_{0}$ and 
\begin{displaymath}
	H := \bigcap^{\infty}_{n=1} \bigcup^{\infty}_{j=n}
	U_{\sigma_{n} (P_{j})} (P_{j}). 
\end{displaymath}
It is easy to see that $H$ is a dense $G_{\delta}$-subset of $\s$.  If
$Q\in H$, then for every $n>0$ there is a $j_{n}$ for which $Q\in
U_{\sigma_{n} (P_{j_{n}})} (P_{j_{n}})$.  Define $D :=
\bigcap^{\infty}_{m=1} \bigcup_{n\geq m} D_{\tau_{n}}^{P_{j_{n}}}(n)
\cap E_{n}$.  By (\ref{directions}), $|D|=1$. This means that, for
each $\vartheta \in D$, there is a subsequence $\{n_{k}\}$ such that
$\vartheta \in D_{\tau_{n_{k}}}^{P_{j_{n_{k}}}} (n_{k})$ for all
$k$. In order to avoid heavy notation, let us denote such a sequence
by $\{n\}$. Now call $C(\vartheta) := \bigcap^{\infty}_{m=1}
\bigcup_{n\geq m} C_{\tau_{n}}^{P_{j_{n}}} (\vartheta,n) \cap G_{n}
(\vartheta)$.  From (\ref{initcond}), $A(C(\vartheta)) =
A(\Su^{Q}(\vartheta))$.

So, for each $\vartheta \in D$ and $x\in C(\vartheta)$ (i.e.,
a.e.~$\vartheta$ and a.e.~$x\in \Su^{Q} (\vartheta)$) there exists a
subsequence $\{ n_{k} \}$ such that $\lim_{k\to \infty} \tau_{n_{k}} =
+\infty$ and
\begin{displaymath}
	\lim_{k\to \infty} \frac{1}{\tau_{n_{k}}}
	\int_{0}^{\tau_{n_{k}}} f_{i}^{Q} \circ \phi^{Q}_{t}(x) dt =
	\frac{1}{A(\Su^{Q})} \int_{\Su^{Q}} f_{i}^{Q} dA,
\end{displaymath}
for all $i\in \N $. Since $\{f_{i}^{Q}\}_{i\in N}$ is dense in $L^{1}
(\Su^{Q}(\vartheta))$, using a standard approximation argument, and
Birkhoff's Theorem, we conclude that $\phi_{\vartheta,t}^{Q}$ is
ergodic for a.e. $\vartheta \in ]0, \pi/2[$.  
\qed 

\proofof{Corollary \ref{gigiCorollary}} Let $F^{N}\subset
{\mathcal{S}}$ be the collection of finite step polygons with $N$
sides. Each $F^{N}$ is a nowhere dense set in ${\mathcal{S}}$, so the
subset of all finite step polygons $F := \bigcup_{N\geq 4}F^{N}$ is a
meager set of ${\mathcal{S}}$.  By Baire's Theorem, $H$ is a set of
second category, therefore it intersects the complement of $F$. In
other words, $H$ contains infinite step polygons. 
\qed

\end{document}